\documentclass[10pt,twocolumn,prb,byrevtex,showpacs]{revtex4}
\usepackage{amsmath}
\usepackage{graphics}

\begin{document}

\title{Simple spin models with non-concave entropies}
\author{Hugo Touchette}
\email{ht@maths.qmul.ac.uk}
\affiliation{School of Mathematical Sciences, Queen Mary,
University of London, London E1 4NS, UK}


\begin{abstract}
Two simple spin models are studied to show that the microcanonical entropy can be a non-concave function of the energy, and that the microcanonical and canonical ensembles can give non-equivalent descriptions of the same system in the thermodynamic limit. The two models are simple variations of the classical paramagnetic spin model of non-interacting spins and are solved as easily as the latter model. 
\end{abstract}


\maketitle

\section{Introduction}

\begin{figure}[b]
\resizebox{3.4in}{!}{\includegraphics{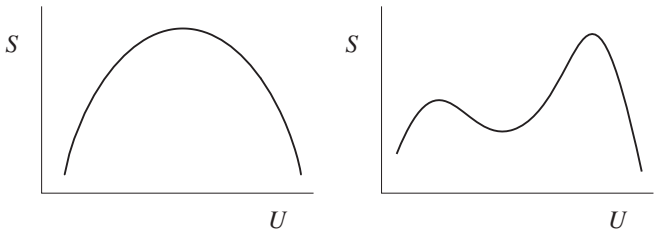}}
\caption{(a) Typical concave entropy $S(U)$. (b) Example of non-concave entropy with two local maxima.}
\label{fignonconcent1}
\end{figure}

The goal of this paper is to present two simple spin models with the property that their entropy $S(U)$ is \emph{non-concave}; that is, $S(U)$ does not have only one maximum, but has several local maxima (Fig.~\ref{fignonconcent1}). It is hoped that these models will interest not only students in physics, but also physicists who have been taught that the entropy is always concave. The fact, as will become clear, is that there is nothing in the definition of the entropy that prevents $S(U)$ from being non-concave. What actually determines the concavity of the entropy is the nature of the interactions and, more precisely, their range. If the components (for example, particles or spins) of a many-body system interact through short-range interactions (as in the nearest-neighbor Ising model), the entropy of such a system is concave in the thermodynamic limit.\cite{note1} However, if the same components interact through long-range interactions (as in mean-field models), then the entropy can be non-concave in the thermodynamic limit.\cite{note2}

Many systems with long-range interactions are known to have non-concave entropies.\cite{draw2002} Examples include systems of gravitating particles used to model stars and galaxies,\cite
{lynden1968,thirring1970,hertel1971,lynden1999,chavanis2006} several spin models,\cite{gross1997,ispolatov2000,barre2001,ellis2004,costeniuc2005a} in addition to statistical models of fluid turbulence,\cite
{ellis2002} which have been used to describe the Great Red Spot of Jupiter\cite{turkington2001} among other interesting phenomena. Most of these models are difficult to solve analytically and numerically, which means that they are not really accessible to students with only a basic knowledge of statistical mechanics.

The two spin models presented here were designed to overcome this problem. The two were constructed so as to provide the simplest proof by example of the fact that the entropy can be non-concave for long-range interactions, and may be presented in an introductory course on statistical mechanics after presenting the ideal spin model of paramagnetism as both are simple variations of that model. The price to pay for this simplicity is that neither of the proposed models is physically realistic. Yet they do well in capturing several interesting features of more realistic models having non-concave entropies, including those mentioned above. Thus they should serve as good toy models for understanding and explaining much of the physics observed in real systems.

Readers who wish to learn more about long-range systems and non-concave entropies are encouraged to read the collection of papers edited by Dauxois et al.~\cite{draw2002} Other accessible references on these subjects will be mentioned at the end of the paper.

\section{\label{sec:2state}Two-state model}

The basic model that serves as a template for the two variations to be studied here is the paramagnetic spin model consisting of $N$ spin-$\frac{1}{2}$ particles $s_1,s_2,\ldots,s_N$ interacting with an external magnetic field $H$.\cite{pathria1996} If we assume that the spins do not interact with each other, we can write the total energy of the system in the usual form
\begin{equation}
U=\mu H\sum_{i=1}^N s_i,
\label{eqh1}
\end{equation}
where $\mu$ is the magnetic moment of the spins, and $s_i$ is the spin variable with the value $-1$ when the $i$th spin is aligned opposite to the magnetic field and $s_i=+1$ when the spin is aligned with the field. 

\begin{figure*}[t]
\includegraphics{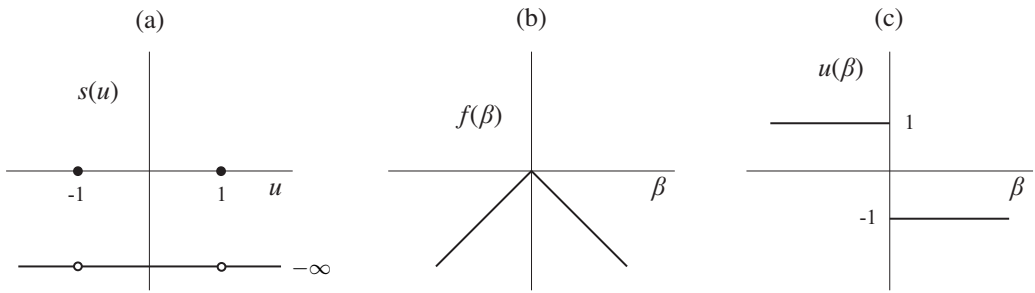}
\caption{Two-state model. (a) Thermodynamic entropy $s(u)$. (b) Thermodynamic free energy $f(\beta)$. (c) Equilibrium energy per spin $u(\beta)=f'(\beta)$ in the canonical ensemble.}
\label{figtwostateent1}
\end{figure*}

The first variation of this model that we consider is constructed by assuming that there is a constraint on the spins that has the effect of forcing them to all take the same state. That constraint is effectively a long-range interaction, because it affects all the spins, and can be thought of as arising from an infinite ferromagnetic interaction acting between the spins or from an infinite energy cost assigned to configurations other than the two completely-aligned configurations.\cite{nagle1968} For our purposes we do not need to specify the nature of the constraint other than just assuming that it is there and that only two possible configurations remain, namely,
\begin{subequations}
\begin{align}
s_1=s_2=\cdots=s_N=-1,\\
\noalign{\noindent and}
s_1=s_2=\cdots=s_N=+1,
\end{align}
\end{subequations}
to which are associated two energy values, $U=-N$ and $U=N$, in units where $\mu H=1$. The entropy associated with the two values of the energy is
\begin{equation}
S(U=-N)=S(U=N)=\ln(1)=0
\end{equation}
in units where $k_B=1$.\cite{leff1999} We can also write $S(U)=\ln(0)=-\infty$ for all $U\neq\pm N$ because there are no spin configurations of energy different than $\pm N$ with the added constraint. Therefore, in the thermodynamic limit, we obtain
\begin{equation}
s(u)=\lim_{N\rightarrow\infty}\frac{S(Nu)}{N}
=
\begin{cases}
0 & \text{if $u=\pm 1$} \\
-\infty & \text{otherwise}
\end{cases}
\label{eqs1}
\end{equation}
for the entropy per spin expressed in terms of the energy per spin $u=U/N$. Equation~\eqref{eqs1} is the thermodynamic entropy of the spin model with the constraint on the spin value.

The free energy of this model is calculated just as easily. The partition function is
\begin{equation}
Z(\beta)=e^{\beta N}+e^{-\beta N},
\end{equation}
because only two energy values are allowed, each associated with only one spin configuration, so that
\begin{equation}
F(\beta)=-\ln Z(\beta) = -\ln(e^{\beta N}+e^{-\beta N}).
\label{free}
\end{equation}
Note that we have omitted the factor of $1/\beta$ in Eq.~\eqref{free}. The free energy per spin $f(\beta)$ is found by by taking the thermodynamic limit:
\begin{equation}
f(\beta)=\lim_{N\rightarrow\infty} \frac{F(\beta)}{N}=-|\beta|.
\label{eqphi1}
\end{equation} 

The functions $s(u)$ and $f(\beta)$ are plotted in Fig.~\ref{figtwostateent1}(a) and \ref{figtwostateent1}(b), respectively. Two properties of these functions are worth noting. The first is that $s(u)$ is a non-concave function of $u$ having two maxima corresponding to its two finite values. The second is that $f(\beta)$ has a jump in its derivative at $\beta=0$. Because the derivative of the free energy of a thermodynamic system yields the equilibrium energy of that system as a function of its temperature, we expect that the discontinuity of the derivative of $f(\beta)$ is related to a phase transition. In this case the phase transition is trivial: it arises because the energy per spin $u$ is discrete and can take only two values. For negative temperatures, the equilibrium energy per spin of the model is $u(\beta)=f'(\beta)=1$ with all the spins pointing in the up direction. As $\beta$ crosses the transition, $\beta=0$, the energy per spin switches to $u(\beta)=-1$ (the block of spins reverses direction), and remains at this value for all positive temperatures because $f'(\beta)=-1$ for $\beta>0$. The complete behavior of $u(\beta)$ is illustrated in Fig.~\ref{figtwostateent1}(c).

\section{Two-block model}

The entropy of the two-state model of Sec.~\ref{sec:2state} is non-concave as a function of the energy per spin, but in a rather singular and contrived way. Is it possible to construct a spin model having an entropy which is non-concave and continuous? The answer is yes. 

\begin{figure*}[t]
\includegraphics{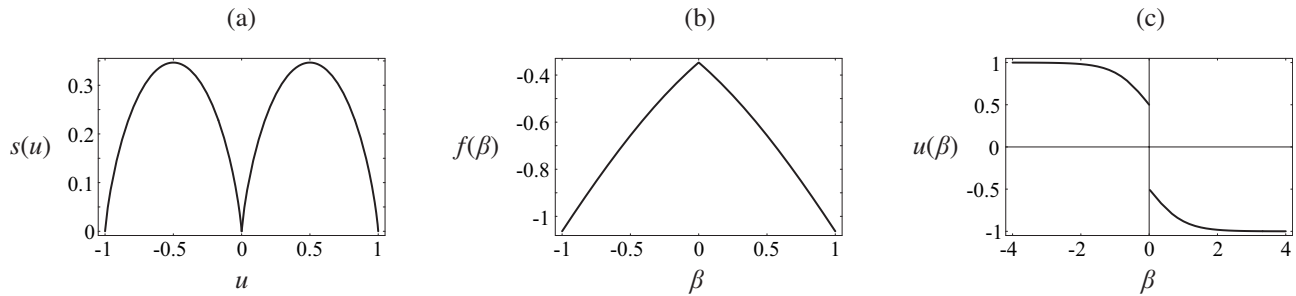}
\caption{Two-block model. (a) Thermodynamic entropy $s(u)$. (b) Thermodynamic free energy $f(\beta)$. (c) Equilibrium energy per spin $u(\beta)=f'(\beta)$ in the canonical ensemble.}
\label{figbsent1}
\end{figure*}

Consider the block of ``frozen'' spins $s_1,s_2,\ldots,s_N$ that we considered earlier, which are constrained to take the same value. We add to this block $N$ new spins $\sigma_1,\sigma_2,\ldots,\sigma_N$ which are free to take the values $\pm 1$ independently of one another. That is, the spins $\sigma_i$ are non-interacting as in the standard paramagnetic spin model. Also the $s$-spins do not interact with the $\sigma$-spins, so that the total energy of the composite system can be written as
\begin{equation}
U=U_s+U_\sigma=\sum_{i=1}^N s_i+\sum_{i=1}^N \sigma_i=Ns_1+\sum_{i=1}^N \sigma_i,
\label{eqh2}
\end{equation}
where $s_i,\,\sigma_i=\pm 1$. (We assume that $\mu H=1$.) The minimum value of the total energy corresponding to the ground state of the system is now $U=-2N$, and its maximum value, corresponding to the excited energy, is $U=2N$. For each of these energies, we have $u=U/(2N)=\pm 1$ as the minimum and maximum energy per spin, respectively.

The thermodynamic entropy and free energy of the two block model can be calculated exactly. The spins $s_i$ and $\sigma_i$ have only one ground state corresponding to all the spins being in the down direction, so that the entropy is $s(-1)=0$. Similarly, 
\begin{equation}
s_1=s_2=\cdots=s_N=\sigma_1=\sigma_2=\cdots=\sigma_N=+1
\end{equation}
is the only spin configuration which realizes the excited energy per spin $u=1$, so that $s(1)=0$. In the middle of the energy spectrum, there are two configurations with $u=0$, which are obtained by putting all the spins $s_i$ in one direction and all the spins $\sigma_i$ in the opposite direction. As a result, we must also have $s(0)=0$. Hence there are three values of $u$ at which $s=0$. Moreover, it is easy to see that $s(u)\neq0$ when $u\neq -1,0,1$, and that $s(u)$ is symmetric about $u=0$. Therefore, $s(u)$ should be a non-concave function of $u$ having two symmetric maxima located between three zeros.

To confirm this guess, let us calculate $s(u)$. Note that the number $\Omega(U)$ of spin configurations of the composite system with energy $U=U_s+U_\sigma$ can be decomposed as
\begin{equation}
\Omega(U)=\Omega_\sigma(U_\sigma=U+N)+\Omega_\sigma(U_\sigma=U-N),
\label{eqotot}
\end{equation}
where $\Omega_\sigma(U_\sigma)$ is the number of configurations of spins $\sigma_i$ with energy $U_\sigma$. These spins form on their own a system of $N$ free spins, the thermodynamic entropy of which is
\begin{equation}
s_\sigma(u)=-\left( \frac{1-u}2\right) \ln \left( \frac{1-u}2\right) -\left( \frac{1+u}2\right) \ln \left( \frac{1+u}2\right), 
\end{equation}
where $u\in[-1,1]$. Therefore, we can write
\begin{equation}
\Omega_\sigma(U_\sigma)\approx e^{Ns_\sigma(U_\sigma/N)},
\end{equation}
with sub-exponential corrections in $N$. If we use this approximation in Eq.~(\ref{eqotot}) and evaluate the limit 
\begin{equation}
s(u)=\lim_{N\rightarrow\infty}\frac{1}{2N}\ln\Omega(2Nu),
\end{equation}
we obtain
\begin{equation}
s(u)=\frac{1}{2}\max\{ s_\sigma(2u-1),s_\sigma(2u+1)\}
\end{equation}
or, equivalently,
\begin{equation}
s(u)=
\begin{cases}
\frac{1}{2}s_\sigma(2u+1) & \text{if $u\in [-1,0[$} \\ 
\frac{1}{2}s_\sigma(2u-1) & \text{if $u\in [0,1]$}.
\end{cases}
\
\label{eqs2}
\end{equation}
The plot in Fig.~\ref{figbsent1}(a) shows that this entropy has the anticipated form. The presence of the two maxima obviously make $s(u)$ non-concave, but unlike our previous example, $s(u)$ is now continuous in $u$. 

To understand the exact form of $s(u)$, note that the spins $s_i$ do not contribute to the entropy of the whole system, as is evident from Eq.~(\ref{eqotot}). The sole function of these spins is to give or absorb two macroscopic amounts of energy, given by $U_s=-N$ and $U_s=N$, respectively. In this sense, the spins $s_i$ can be thought as playing the role of a reservoir of energy or an ``energy switch.'' When $U_s=-N$, the total energy $U$ must be in the range $[-2N,0]$, so that $s(u)$ for $u\in[-1,0]$ is the entropy of the spins $\sigma_i$ with 
\begin{equation}
u_\sigma=\frac{U_\sigma}{N}= \frac{U+N}{N}=2u+1
\end{equation}
as the energy per spin of the spins $\sigma_i$. Similarly, when $U_s=N$, $U\in[0,2N]$, so that $s(u)$ for $u\in[0,1]$ corresponds to the entropy of the spins $\sigma_i$, the energy per spin of which is $u_\sigma=2u-1$. The extra factor $1/2$ in front of $s_\sigma$ is present because the entropy per spin of the composite system is calculated for $2N$ spins, not just $N$.

We next calculate $f(\beta)$ by first evaluating the partition function
\begin{subequations}
\begin{align}
Z(\beta)	&=\sum_{s_1\pm 1,\sigma_1,\sigma_2,\ldots,\sigma_N} e^{-\beta(U_s+U_\sigma)}\\
		&=\sum_{s=\pm 1} e^{-\beta Ns}\left(\sum_{\sigma=\pm 1} e^{-\beta\sigma}\right)^N\\
		&= (e^{\beta N}+e^{-\beta N})(e^{\beta}+e^{-\beta})^N.
\end{align}
\end{subequations}
By taking the limit $N\rightarrow\infty$ as in Eq.~(\ref{eqphi1}), but with $N$ replaced by $2N$, we obtain
\begin{equation}
f(\beta)=-\frac{|\beta|+\ln (2\cosh\beta)}{2}.
\label{eqfr2}
\end{equation}

This result is plotted in Fig.~\ref{figbsent1}(b). As for the previous model, $f(\beta)$ is non-differentiable at $\beta=0$ because the term in $| \beta |$, which corresponds to the free energy of the frozen spins $s_i$, is non-differentiable at this point. The phase transition originating from the discontinuity of $f'(\beta)=u(\beta)$ is illustrated in Fig.~\ref{figbsent1}(c), and can be interpreted as before by noting that the block of spins $s_i$ changes orientation when $\beta$ goes continuously from $\beta=0^-$ to $\beta=0^+$.\cite{note3} 

\section{Non-equivalent ensembles}

By now there should be no doubt that $s(u)$ can be a non-concave function. What does it mean physically for $s(u)$ to be non-concave? If we are to apply the common interpretation of the temperature as given by
\begin{equation}
\beta=\frac{1}{k_B T}=s'(u),
\label{eqes1}
\end{equation}
then it seems as if the temperature of a system with a non-concave entropy varies non-monotonically as a function of the internal energy. This interpretation of Eq.~(\ref{eqes1}) is valid if we view $\beta$ as nothing but the derivative of the entropy. However, if we follow the definition of the canonical ensemble and view $\beta$ as a parameter fixed externally by a heat reservoir, then there is a problem. In this case, the equilibrium value $u(\beta)$ of the energy per spin, which is normally determined by solving Eq.~(\ref{eqes1}), must be a multi-valued function of $\beta$ if $s(u)$ is non-concave, which is not what we see in Fig.~\ref{figbsent1}(c). Indeed, apart from the point $\beta=0$, $u(\beta)$ takes only one value for each $\beta$ even though $s(u)$ is non-concave. Thus it seems that for a non-concave entropy, the determination of the equilibrium energy from $\beta=s'(u)$ or from the derivative of $f(\beta)$ yields two different results. What is wrong?

What is wrong, in a nutshell, is that the fundamental relation (\ref{eqes1}) does not only determine the equilibrium value $u(\beta)$ in the canonical ensemble when $s(u)$ is non-concave; it also determines a metastable value for $u$ as well as an unstable value. What is seen by taking the derivative of $f(\beta)$ is only the stable root of Eq.~(\ref{eqes1}) corresponding to the value of $u$ realized at equilibrium in the canonical ensemble.

This explanation may be difficult for students to follow in a first course on equilibrium statistical mechanics (however, see Problem 3). A much more transparent and related consequence of the non-concavity of the entropy which can be explained to them is that the thermodynamic description provided by the microcanonical ensemble as a function of the energy ceases to be completely compatible with the thermodynamic description provided by the canonical ensemble as a function of the temperature. In short, the two ensembles cease to be equivalent.

The two-block model can be used to illustrate this point. If we go back to the behavior of $u(\beta)$ observed for this model (Fig.~\ref{figbsent1}(c)), we see that $u(\beta)$ never takes any value in the range $(-\frac{1}{2},\frac{1}{2})$ because of the discontinuity at $\beta=0$. In the microcanonical ensemble, the energy per spin $u$ can be ``tuned'' anywhere in the range $[-1,1]$, and to each $u\in[-1,1]$ is associated a well-defined value of the entropy $s(u)$ (Fig.~\ref{figbsent1}(a)). The two ensembles must therefore be non-equivalent because there are values of $u$ that can be realized in the microcanonical ensemble, but not in the canonical ensemble. In the present case, the equilibrium states of the microcanonical ensemble related to $u\in (-\frac{1}{2},\frac{1}{2})$ have no equivalent equilibrium states in the canonical ensemble because these values of $u$ do not show up at equilibrium in the latter ensemble for any values of $\beta$; compare Figs.~\ref{figbsent1}(a) and \ref{figbsent1}(c).

Another approach for demonstrating the non-equivalence of the microcanonical and canonical ensembles is to show that the Legendre transform of $f(\beta)$ does not yield the complete entropy $s(u)$ when the latter is non-concave.\cite{note4} Although more mathematical in nature, this approach works well for the two-block model because the Legendre transform of the two differentiable branches of $f(\beta)$ can be calculated directly for this model. The result of this calculation\cite{note5} is that the Legendre transform of $f(\beta)$ yields the correct expression of $s(u)$ given in Eq.~(\ref{eqs2}) for all $u\in[-1,1]$ except $u\in(-\frac{1}{2},\frac{1}{2})$, which is exactly the range of values of $u$ not covered by $u(\beta)$ and the region of non-equivalent ensembles discussed before. 

\section{Outlook and suggested problems}

The two-state and two-block models can be used to explore many other intriguing properties of systems with non-concave entropies. Provided below are six problems covering some of these properties. The solutions of these problems can be found in an extended version of this paper,\cite{touchette2005c} in addition to two short introductions to the theory of non-concave entropies and non-equivalent ensembles.\cite{touchette2004b,touchette2005a} For more information about the development of this young theory, the reader is invited to read the author's thesis\cite{touchette2003} and the papers of Thirring,\cite{{thirring1970}} Hertel,\cite{hertel1971} and Lynden-Bell.\cite{lynden1999}

\emph{Problem 1}. Show that the Legendre transform of $s(u)$ for either the two-state or the two-block model yields the correct $f(\beta)$ for all $\beta$. There is a difficulty here: because $s(u)$ is non-concave for both models, how should the Legendre transform be defined? The answer is
\begin{equation}
f(\beta)=\min_u \{\beta u -s(u)\},
\end{equation}
where ``min'' stands for ``minimum of.'' This transform is called a \emph{Legendre-Fenchel} transform.

\emph{Problem 2}. Prove in general that if $s(u)$ is non-concave, then $f(\beta)$ must have one or more non-differentiable points. In other words, prove that a non-concave entropy in the microcanonical ensemble implies a discontinuous or first-order phase transition in the canonical ensemble. 

\emph{Problem 3}. Show that the distribution of the energy per particle $u=U/N$ in the canonical ensemble, given by
\begin{equation}
P_\beta(u)=\frac{\Omega(Nu) e^{-\beta Nu}}{Z(\beta)}\approx e^{-N[\beta u -s(u)-f(\beta)]},
\end{equation}
has more than one maximum when $s(u)$ is non-concave. Show also that the positions of the various maxima of $P_\beta(u)$ are given for $N\rightarrow\infty$ by Eq.~(\ref{eqes1}). Which of these maxima corresponds to the equilibrium value of $u$? Which is the metastable value? Does $P_\beta(u)$ have more than one maximum when $s(u)$ is concave?

\emph{Problem 4}. Show that a system with a non-concave entropy can have values of $u$ at which its heat capacity, calculated in the microcanonical ensemble for fixed values of $u$, is negative. Can the heat capacity be negative in the canonical ensemble? 

\emph{Problem 5}. Show that you can recover the full non-concave entropy of the two-state or two-block model by calculating the Legendre transform of the Gaussian free energy $f_\gamma(\beta)$ associated with the following Gaussian partition function:
\begin{equation}
Z_\gamma(\beta)=\sum_{\rm microstates} e^{-\beta U-\gamma U^2}.
\end{equation}
How should $\gamma$ be chosen? [The calculation of $Z_\gamma(\beta)$ for a model similar to the two-state model can be found in Ref.~\onlinecite{touchette2006a}. The theory of Gaussian free energies and their relation to non-concave entropies can be found in Ref.~\onlinecite{costeniuc2006}.]

\emph{Problem 6}. Given that the microcanonical and canonical ensembles are non-equivalent for systems with non-concave entropies, under which conditions or principles should we use one ensemble or the other to describe such systems? Is one of the two ensembles more ``physical'' or more ``natural'' than the other?

\begin{acknowledgments}
I would like to thank W.\ Krauth for his many constructive comments on this paper, which was (re)written while holding a grant from HEFCE (England).
\end{acknowledgments}


\begin{thebibliography}{1}

\bibitem{note1} Precise mathematical results expressing this statement can be found in G. Gallavotti, \textit{Statistical Mechanics: A Short Treatise} (Springer, New York, 1999), Chap.~4 or O. E. Lanford, ``Entropy and equilibrium states in classical statistical mechanics,'' in \textit{Statistical Mechanics and Mathematical Problems}, edited by A. Lenard (Springer, Berlin, 1973), pp.~1--113, Sec.~A.

\bibitem{note2} Systems with short-range interactions have concave entropies because they are \emph{additive}, that is, they can be decomposed into non-interacting parts in the thermodynamic limit. Systems with long-range interactions are not additive; hence the fact that they can have non-concave entropies. See the first chapter of Ref.~\onlinecite{draw2002} for a good exposition of this reasoning.

\bibitem{draw2002} \textit{Dynamics and Thermodynamics of Systems with Long Range Interactions}, edited by T. Dauxois, S. Ruffo, E. Arimondo, and M. Wilkens (Springer, New York, 2002). 

\bibitem{lynden1968} D. Lynden-Bell and R. Wood, ``The gravo-thermal catastrophe in isothermal spheres and the onset of red-giant structure for stellar systems,'' Mon. Notic. Roy. Astron. Soc. \textbf{138}, 495--525 (1968). 

\bibitem{thirring1970} W. Thirring, ``Systems with negative specific heat,'' Z. Physik A\textbf{235}, 339--352 (1970). 

\bibitem{hertel1971} P. Hertel and W. Thirring, ``A soluble model for a system with negative specific heat,'' Ann. Phys. (NY) \textbf{63}, 520--533 (1971). 

\bibitem{lynden1999} D. Lynden-Bell, ``Negative specific heat in astronomy, physics and chemistry,'' Physica A \textbf{263}, 293--304 (1999). 

\bibitem{chavanis2006} P.-H. Chavanis, ``Phase transitions in self-gravitating systems,'' Int. J. Mod. Phys. B \textbf{20}, 3113--3198 (2006). 

\bibitem{gross1997} D. H. E. Gross, ``Microcanonical thermodynamics and statistical fragmentation of dissipative systems: The topological structure of the $N$-body phase space,'' Phys. Rep. \textbf{279}, 119--202 (1997). 

\bibitem{ispolatov2000} I. Ispolatov and E. G. D. Cohen, ``On first-order phase transitions in microcanonical and canonical non-extensive systems,'' Physica A \textbf{295}, 475--487 (2000). 

\bibitem{barre2001}J. Barr\'e, D. Mukamel, and S. Ruffo, ``Inequivalence of ensembles in a system with long-range interactions,'' Phys. Rev. Lett. \textbf{87}, 030601-1--4 (2001). 

\bibitem{ellis2004} R. S. Ellis, H. Touchette, and B. Turkington, ``Thermodynamic versus statistical nonequivalence of ensembles for the mean-field Blume-Emery-Griffiths model,'' Physica A \textbf{335}, 518--538 (2004).

\bibitem{costeniuc2005a} M. Costeniuc, R. S. Ellis, and H. Touchette, ``Complete analysis of phase transitions and ensemble equivalence for the Curie-Weiss-Potts model,'' J. Math. Phys. \textbf{46}, 063301-1--25 (2005).

\bibitem{ellis2002} R. S. Ellis, K. Haven, and B. Turkington, ``Nonequivalent statistical equilibrium ensembles and refined stability theorems for most probable flows,'' Nonlinearity \textbf{15}, 239--255 (2002).

\bibitem{turkington2001} B. Turkington, A. Majda, K. Haven, and M. DiBattista, ``Statistical equilibrium predictions of jets and spots on Jupiter,'' Proc. Nat. Acad. Sci. (USA) \textbf{98}, 12346--12350 (2001). 

\bibitem{pathria1996} See, for example, R. K. Pathria, \textit{Statistical Mechanics} (Butterworth Heinemann, Oxford, 1996), 2nd ed., Sec.~3.9.

\bibitem{nagle1968} J. F. Nagle, ``The one-dimensional KDP model in statistical mechanics,'' Am. J. Phys. \textbf{36}, 1114--1117 (1968).

\bibitem{leff1999} H. S. Leff, ``What if entropy were dimensionless?,'' Am. J. Phys. \textbf{67}, 1114--1122 (1999).

\bibitem{note3} Other examples of one-dimensional models having phase transitions in the canonical ensemble can be found in C. Kittel, ``Phase transition of a molecular zipper,'' Am. J. Phys. \textbf{37}, 917--920 (1969) and Ref.~\onlinecite{nagle1968}. 

\bibitem{note4} Recall that the Legendre transform of $f(\beta)$ yields $s(u)$ when the latter is concave.

\bibitem{note5} See Refs.~\onlinecite{touchette2005c} and \onlinecite{touchette2003} for the complete calculation.

\bibitem{touchette2005c} H. Touchette, ``A simple spin model with nonequivalent microcanonical and canonical ensembles,'' cond-mat/0504020v1 (2005). 

\bibitem{touchette2004b} H. Touchette, R. S. Ellis, and B. Turkington, ``An introduction to the thermodynamic and macrostate levels of nonequivalent ensembles,'' Physica A \textbf{340}, 138--146 (2004). 

\bibitem{touchette2005a} H. Touchette and R. S. Ellis, ``Nonequivalent ensembles and metastability,'' in \textit{Complexity, Metastability and Nonextensivity}, edited by C. Beck, G. Benedek, A. Rapisarda, and C. Tsallis (World Scientific, Singapore, 2005), p.~81.

\bibitem{touchette2003} H. Touchette, ``Equivalence and nonequivalence of the microcanonical and canonical ensembles: A large deviations study,'' Ph.D. thesis, Department of Physics, McGill University (2003). Available at \url{<www.maths.qmul.ac.uk/~ht>}.

\bibitem{touchette2006a} H. Touchette and C. Beck, ``Nonconcave entropies in multifractals and the thermodynamic formalism,'' J. Stat. Phys. \textbf{125}, 455--471 (2006). 

\bibitem{costeniuc2006} M. Costeniuc, R. S. Ellis, H. Touchette, and B. Turkington, ``Generalized canonical ensembles and ensemble equivalence,'' Phys. Rev. E \textbf{73}, 026105-1--8 (2006). 

\end{thebibliography}
\end{document}